\documentclass[11pt]{article} % use larger type; default would be 10pt

\usepackage[utf8]{inputenc} % set input encoding (not needed with XeLaTeX)

\usepackage{geometry} % to change the page dimensions
\usepackage{graphicx} % support the \includegraphics command and options

\usepackage{booktabs} % for much better looking tables
\usepackage{array} % for better arrays (eg matrices) in maths
\usepackage{paralist} % very flexible & customisable lists (eg. enumerate/itemize, etc.)
\usepackage{verbatim} % adds environment for commenting out blocks of text & for better verbatim
\usepackage{subfig} % make it possible to include more than one captioned figure/table in a single float
\usepackage{dsfont}

\usepackage{fancyhdr} % This should be set AFTER setting up the page geometry
\usepackage{sectsty}
\usepackage[nottoc,notlof,notlot]{tocbibind} % Put the bibliography in the ToC
\usepackage[titles,subfigure]{tocloft} % Alter the style of the Table of Contents

\usepackage{authblk}
\usepackage{hyperref}
\usepackage{cite}
\title{Benchmarking Framework for Performance-Evaluation of Causal Inference Analysis}
\author[1]{Yishai Shimoni\thanks{yishais@il.ibm.com}}
\author[1]{Chen Yanover\thanks{cheny@il.ibm.com}}
\author[1]{Ehud Karavani}
\author[1]{Yaara Goldschmnidt}
\affil[1]{IBM Research - Haifa}
\begin{document}
\maketitle

\begin{abstract}
Causal inference analysis is the estimation of the effects of actions on outcomes.
In the context of healthcare data this means estimating the outcome of counter-factual treatments
(i.e. including treatments that were not observed) on a patient's outcome. 
Compared to classic machine learning methods, 
evaluation and validation of causal inference analysis is more challenging
because ground truth data of counter-factual outcome can never be obtained in any real-world scenario.
Here, we present a comprehensive framework for benchmarking algorithms that estimate causal effect.
The framework includes unlabeled data for prediction, 
labeled data for validation, 
and code for automatic evaluation of algorithm predictions using both established and novel metrics.
The data is based on real-world covariates, and the treatment assignments and outcomes are based on simulations,
which provides the basis for validation.
In this framework we address two questions: 
one of scaling, and the other of data-censoring.
The framework is available as open source code at 
\href{https://github.com/IBM-HRL-MLHLS/IBM-Causal-Inference-Benchmarking-Framework}
{https://github.com/IBM-HRL-MLHLS/IBM-Causal-Inference-Benchmarking-Framework}
\end{abstract}

\section{Introduction}
Estimating the causal effect of an intervention on some measurable outcome
(e.g. the effect of a treatment or procedure on survival) is a vibrant field of research.
It has gained much traction in the last decade with the rise of big-data 
\cite{winship1999estimation,Hernan2016Emulate}, especially in the context of health-care 
(e.g., see \cite{schneeweiss2009high, dormuth2014higher, madigan2014systematic, gottlieb2017estimating}).
Effect estimation from real-world observational data requires estimating all \emph{counter-factual outcomes}
of a treatment
(e.g. the outcome with and without treatment).
This estimation is complicated since for each individual 
we can only observe the factual outcome associated with the treatment that was actually assigned to the individual.

One way of tackling this problem is through randomized controlled trials (RCTs).
By randomly assigning individuals to treatment groups
it is possible to consider the average outcome in each such treatment group 
as an unbiased estimate of the outcome in the entire population. 
This allows calculating the average treatment effect in the population. 
However, RCTs are labor- and time-intensive and, in some cases, 
may even by unethical or unfeasible. 
Additionally, the population that is being recruited to an RCT is almost always pre-selected 
and does not represent the global population. 
Finally, RCT-based estimates may also be biased, 
for example when informative censoring occurs
\cite{concato2000randomized}.

Estimating treatment effect from real-world observational data offers several advantages 
and several drawbacks as compared to RCTs. 
On the one hand, such data represents much larger and less homogeneous populations,
and thus estimates based on such data are more immediately applicable to the general population 
and may also be available for rare diseases.
On the other hand, assignment of treatment typically depends on individual characteristics 
(e.g., the severity of the disease), 
some of which may also affect the outcome of that same treatment, 
and therefore introduce (confounding) bias. 
Correcting for such biases is required for an accurate estimation of treatment effect. 
Indeed, many different methods have been devised for 
and applied to this task, including inverse probability weighting (IPW) 
\cite{austin2011introduction}, 
g-formula \cite{keil2014parametric}, 
and doubly robust methods \cite{robins2007comment}.

Despite great advances in causal inference research, 
it is still unclear which algorithm should be used for a given effect estimation instance.
Moreover, it is unclear what data characteristics (e.g. number of covariates or samples) 
need to be considered when selecting an algorithm and how this affects performance. 
In part, this is due to a lack of common data sets and evaluation measures that are 
used as agreed-upon benchmarks. 

Many sub-fields of machine-learning have established
benchmarking datasets that allow comparisons between methods.
Such datasets exist, for example, for
handwriting recognition \cite{lecun1998gradient},
object detection \cite{deng2009imagenet},
and sentiment analysis in natural language processing \cite{go2009twitter},
among others.
However, no such benchmarking dataset exists for causal inference analysis.

Here we present a comprehensive framework 
for offline benchmarking of methods for causal effect inference,
called the \emph{IBM Causal Inference Benchmarking Framework},
which is available online as open-source code. 
The framework contains a python package that allows evaluation of the estimated effects 
by providing several non-redundant scores.
This evaluation can be applied to any given set of ground-truth data and inference data.
Specifically, the framework includes a dataset, as detailed below, 
that can allow a standardized comparison using a 
unified dataset and a unified evaluation code.

Since individual ground-truth data of causal effect can never be known 
for any real-world treatment, 
we developed a simulation-based approach that uses a set of covariates 
and creates a causal graph to determine treatment assignment and effect. 
The framework contains multiple pairs of simulated treatment assignments and effects, 
and their associated ground-truth counter-factual data
(i.e. the outcome for each value of the treatment assignment). 
To ensure that the evaluated performance has real-world implications
we used a cohort of 100K samples derived from 
the publicly available Linked Births and Infant Deaths Database (LBIDD)
\cite{macdorman1998infant} as the fundamental set of covariates.

The data is divided into two sets, each aimed to answer a different question
related to large real-world observational data. 
The first question is one of scaling - can a method take advantage of increasing data sizes, 
and at what computational cost? 
To answer this question, we include data sub-sets comprised of files with varying sizes. 
The performance on each size can be evaluated separately and a global score can be calculated, 
summarizing the performance of the method on the whole data set. 
The second question is regarding informative censoring - 
what methods perform better when outcome is missing for a non-random subset of the samples? 
Such a scenario may arise, for example, 
when a treatment is very effective and as a result there is no follow-up, 
or when a treatment has severe side-effects 
and therefore for some of the individuals we do not observe the results of the treatment. 
To address this question, 
we include data sets in which some of the outcomes are replaced by an {\tt NA} value
according to some underlying pre-determined model (built on top of the features).

In both sets, each simulation file is associated with meta-data indicating the type of simulation that generated it.
This includes information such as the number of covariates, 
the number of confounding covariates (that affect both treatment assignment and outcome), 
and the levels of non-linearity.
This information can be used to further analyze the variables affecting performance.

The scoring code is in Python (versions 2.7 or $\ge 3.4$) and is publicly available 
as an open-source (Apache 2.0 license) github repository at
\href{https://github.com/IBM-HRL-MLHLS/IBM-Causality-Benchmarking-Framework}
{https://github.com/IBM-HRL-MLHLS/IBM-Causality-Benchmarking-Framework}.
A full usage description is available in the code repository.

\section{Methods}

\subsection{Notation}
There are several ways to define causal effect \cite{hernan2018causal}.
Let us define the counter-factual outcome for individual $i$ 
as the outcome that might be observed had they received treatment $a$, 
denoted by $y^a_i$. 
In the context of this benchmarking platform we use the additive treatment effect, 
which is the difference between the treated and untreated counter-factual outcomes.
Therefore, the \emph{individual treatment effect} for individual $i$
is defined as $y^1_i - y^0_i$, where $a=1$ denotes treatment and $a=0$ denotes no-treatment.
The average of the individual treatment effect across the whole population is defined as
the \emph{population treatment effect}, 
or average treatment effect (ATE) \cite{Holland1978-statistics}.

\subsection{Data Description}
The data is conceptually comprised of three main components:
\begin{itemize}
\item{Covariate table: A table holding features in the columns and samples 
(i.e. individuals) in rows, 
and serves as the basis for all the simulated observed outcomes.
The data is based on real-world clinical measurements taken from the Linked Birth and Infant Death Data (LBIDD) 
\cite{macdorman1998infant}.}
\item{Factual outcome files (observation files): A set of three-column files containing sample id,
simulated treatment assignment, and simulated observed outcome. 
This emulates the clinical data that 
may be available in real-world observational databases.}
\item{Counter-factual outcome files (label files): A set of files containing sample id, 
simulated outcome without treatment, and simulated outcome with treatment.
These files serve as the labeled data and used for evaluating the prediction results.}
\end{itemize}

\begin{figure}[t]
\includegraphics[width=\columnwidth]{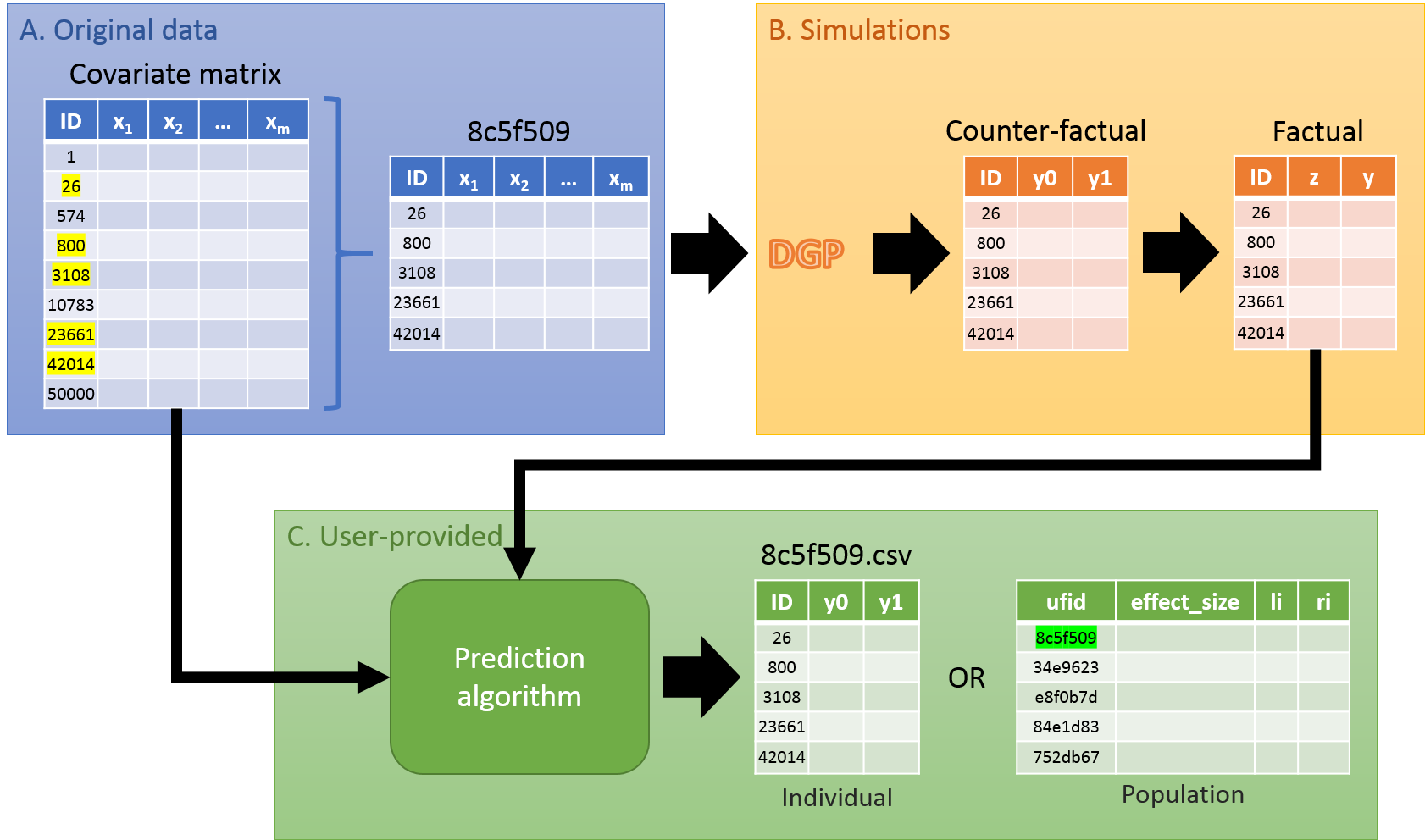}
\caption{
A schematic of the data-flow in the benchmarking platform. 
(A) Covariate data is provided and a subset of samples is randomly selected. 
(B) A data generating process (DGP) is applied using the selected subset 
in order to create the files containing the counter-factual outcomes (label files).
The observation files are created 
by only selecting the relevant counter-factual outcome
(or providing an {\tt NA} value when censored).
(C) The prediction algorithm can be trained using pairs of observation and label files.
For testing and evaluation, the prediction algorithm uses the observation files 
to create effect prediction estimations, 
which are compared with the label files to obtain the scoring metrics.
}
\label{fig:data}
\end{figure}

The covariate file is named \texttt{x.csv}, 
and is a comma-delimited-file with a header as its first row holding the names of the features.
The covariate file is used to create the observation and label files 
as explained below.
Each sample (row) in the covariates file has a unique 
\emph{sample\_id} to match the corresponding sample id in the observation data
and in the label files
(since these files contain fewer samples than available in the covariate file).

A schematic of the data-flow is provided in Figure~\ref{fig:data}, 
describing the following steps.
First, a data generating process (DGP) is defined to generate 
simulated counter-factual outcome values from the covariate file.
The DGP is defined by a collection of parameters that determine the complexity 
of the causal graph, as well as the functions defining the functional 
relation between the covariates and the outcome, as explained in Section~\ref{sec:DGP}.

Once a DGP is defined, a specific instance of the DGP is used 
(i.e. a model is created that adheres to the comlpexity parameters above) 
to create counter factual outcomes, from which the label file and the observations file are created. 
Such a pair is denoted as a \emph{data instance}.
Each DGP is used multiple time to create DGP instances, 
and each one create one data instance,
where each data instance is based on a sub-group of samples from the covariates file.

The data is conceptually split into two evaluation {\em tracks}: 
one for the evaluation of performance as a function of size ({\tt scaling} directory),
and one for the evaluation of performance in the presence of censored data ({\tt censoring} directory).
In the scaling evaluation track there are files containing $n$ samples each, 
where $n \in \{1k, 2.5k, 5k, 10k, 25k, 50k\}$.
In the censoring evaluation track each file contains $10k$ samples.

In order to discourage reverse-engineering (e.g. to determine which data comes from which DGP), 
each file pair has a unique file identifier (ufid) $u$ (e.g. 8c5f509) 
as its base name which is unrelated to the DGP.
The factual files are named using the base name followed by a {\tt .csv} extension 
(e.g. {\tt 8c5f509.csv}), and are comma-delimited files (with a header) containing the
\emph{sample\_id}, $z$, and $y$,
where $z$ is the treatment assignment and $y$ is the observed outcome 
({\tt NA} is used as the outcome in case of censoring).
The counter-factual files are named using $u$ followed by {\tt \_cf.csv}
suffix (e.g. {\tt 8c5f509\_cf.csv}),
and are also comma-delimited files (with a header) containing the
\emph{sample\_id}, $y^0$, and $y^1$.

\subsection{Output Data Description}
Regardless of the evaluation track (i.e. scaling or censoring), 
results can be evaluated either for \emph{population effect prediction} or \emph{individual effect prediction}.

Population effect prediction requires estimating three values per observed data file: 
(a) the effect size over the population, 
(b) the left boundary of a 95\% confidence interval, and 
(c) the right boundary of a 95\% confidence interval.
Therefore, all estimates for all instances of provided observed data 
can fit into one comma-delimited file 
where each row represents a different data instance $u$ indicated in the first column,
followed by one column for each estimated value above.
The evaluation code assumes a header line (holding {\tt ufid}, {\tt effect\_size}, {\tt li}, {\tt ri}), 
and does not impose any restrictions on the order of data instances used.

Evaluation of individual effect requires two estimates \emph{for each individual} $i$ 
in each observed data instance $u$: 
the predicted outcome without treatment ($y^0_i$), 
and the predicted outcome with treatment ($y^1_i$). 
Consequently, for each data instance $u$, 
there should be a corresponding comma-separated file named {\tt u.csv}. 
In this file, each row corresponds to an individual from the observed data instance 
(by matching the $sample\_id$ key). 
Each row has three columns: {\tt sample\_id} of the same individuals from the observation file,
and two other columns, namely {\tt y0} and {\tt y1}, for the estimated counter-factual outcomes $y^0$ and $y^1$,  respectively.
All these comma-separated files should be placed in a single directory in order for the evaluation script to evaluate them all. 
Please refer to the green panel in Fig \ref{fig:data}C.

\subsection{Scoring}
The benchmarking framework provides several non-redundant scoring metrics that together allow a 
comprehensive and comparative estimation of the performance of each method.

Metrics are calculated separately for each given size of dataset and are later aggregated.
It is possible to provide only a subset of the sizes as input to the scoring code
(e.g. if the running time becomes prohibitive),
and obtain scores for the submitted sizes only.

Let $n \in \{1k, 2.5k, 5k, 10k, 25k, 50k\}$ be the number of samples in a data instance.
We denote $D_n$ to be the set of all the data instances consisting of $n$ samples.
let the true effect size be $e$  and the estimated effect size $\hat{e}$. 
The main score is based on the accuracy of the point-estimated effect.
For the population-effect predictions, where the 95\% confidence interval (CI) is provided, 
several additional metrics based on the CI are included to evaluate the precision of methods as well.

%Let $\hat{l_i}$ be the left boundary of a CI, and $\hat{r_i}$ be the right one.

\subsubsection{ENoRMSE Score}
Accuracy is measured on the estimated effect size using effect-normalized root mean squared error (ENoRMSE). 
This is done since the size of the error, as measured by the bias $\hat{e} - e$, 
depends on the actual effect
(i.e. a large bias for a large effect should count the same as a small bias for a small effect).
The ENoRMSE score for the population effect predictions for a specific $D_n$ is defined as
\begin{equation}
\label{eq:enormse_pop}
E_n = \sqrt{\frac{1}{|D_n|} \sum_{j \in D_n}(1-\frac{\hat{e_j}}{e_j})^2},
\end{equation}
where we add a stabilization constant of $\delta=10^{-7}$ in both numerator and denominator
to avoid division by zero errors.

In a similar way, for individual effect prediction, 
the ENoRMSE score is calculated in the following way.
For each individual $i\in \{1,...,n\} $ in a data instance $j$ we first obtain the estimated individual effect,
$\hat{e_{i,j}} = \hat{Y}^1_{i,j} - \hat{Y}^0_{i,j} $.
We then obtain the true individual effect from the simulated counter-factual values
$e_{i,j} = Y^1_{i,j} - Y^0_{i,j} $.
The ENoRMSE accuracy metric is then defined as
\begin{equation}
E_n = \sqrt{\frac{1}{|D_n|} \sum_{j \in D_n} \frac{1}{n} \sum_{i=1}^n 
(1 - \frac{\hat{e_{i,j}}}{e_{i,j}})^2},
\end{equation}
where here, too, we add a stabilization factor $\delta$ in both parts of the fraction.

\paragraph{RMSE Score}
Since ENoRMSE might be sensitive to small effect-sizes and since 
computational precision tends to decrease near zero, 
the ENoRMSE score might over-penalize estimations for cases with no effect.
We therefore add the framework the standard (a.k.a vanilla) un-normalized root mean square error calculation:
\begin{equation}
\label{eq:rmse_pop}
RMSE_n = \sqrt{\frac{1}{|D_n|} \sum_{j \in D_n}(\hat{e_j} - {e_j})^2}.
\end{equation}

\subsubsection{Bias}
The above accuracy measures square the estimation error, leaving no information regarding the direction of error.
Bias measures the difference between the estimated effect and the true one.
Allowing us to later examine whether certain algorithms tend to over-estimate or under-estimate causal effect.
\begin{equation}
\label{eq:bias_pop}
B_n = \frac{1}{|D_n|} \sum_{j \in D_n}(\hat{e_j} - {e_j}).
\end{equation}

\subsubsection{Coverage}
Coverage is calculated as the percentage of times the estimated confidence interval (CI) of the effect size covered the true effect size,
and is defined by
\begin{equation}
C_n = \frac{1}{|D_n|}\sum_{j\in D_n} \mathds{1}_{[l_j \le e_j \le r_j]},
\end{equation}
where $\mathds{1}_A$ is the indicator function (i.e. equals to 1 if $x$ is true and 0 otherwise), 
and $l_j$ and $r_j$ are the left and right edges of the CI, respectively.

Coverage is a non-parametric estimation of the CIs. 
It offers an intuitive measure of the reliability of the method, 
but only if the CIs are calculated correctly and genuinely.
For example, if a method grossly underestimates its CI then its coverage 
will not be comparable with another method that overestimates its CI.
Additionally, coverage is also prone to misleading results and to hacking, 
(e.g. providing CIs that are $ (-\infty, \infty)$ guarantees full coverage,
but this result is non-informative).

\subsubsection{CIC}
Confidence-interval credibility (CIC) estimates the reliability of the CI.
It is defined as 
\begin{equation}
\xi_n = \frac{1}{|D_n|}\sum_{j\in D_n} \frac{|\hat{e_j} - e_j|}{r_j - l_j}.
\end{equation}

The CIC provides an estimate of the percentage of times when 
the real value is within the confidence interval.
It captures the average ratio between the CI and the actual error.
In general, if the CI reliably calculates the 95\% CI, then this value should approach $0.95$
(except for some pathological cases).
As a result, if the CIC is too far from unity then the CI is unreliable, and the coverage should be ignored.

\subsubsection{ENCIS}
The effect normalized confidence interval size (ENCIS) measures the effective size of the 95\% confidence interval. 
The smaller the CI size -  the more precise the estimation is.
However, as with the ENoRMSE score, 
we normalize by the effect size since larger CI can be tolerated for large effect sizes. 
The ENCIS is defined as

\begin{equation}
\epsilon_n = \frac{1}{|D_n|}\sum_{j\in D_n} \frac{r_j - l_j}{|e_j|}.
\end{equation}
Here too, like in the ENoRMSE score, we add a stabilization term in the numerator and denominator.

\subsubsection{Aggregating Scores}
In the scaling evaluation track scores are obtained for each dataset size 
$n \in \{1k$, $2.5k$, $5k$, $10k$, $25k$, $50k\} $
individually, as described above.
To obtain a single score all the individual scores are aggregated  
so that each sample carries an equal weight.
This implies that larger datasets carry a larger weight in the overall aggregated score,
as desired.

Let $A_n$ be a squared accuracy score (i.e. ENoRMSE and RMSE) for given Datasets of size $n$
as defined in Eq.~\ref{eq:enormse_pop} and Eq.~\ref{eq:rmse_pop}. 
The aggregated score is a weighted quadratic sum of the form
\begin{equation}
E = \sqrt{\frac{1}{\sum_{n} n|D_n|} \cdot \sum_{n}n|D_n|\cdot E_n^2 }.
\end{equation}

For all other precision scores $P_n$ (i.e. the Coverage, CIC, ENCIS and the non-squared accuracy score Bias)
the aggregated score is a weighted sum of the form 
\begin{equation}
P = \frac{1}{\sum_{n}n|D_n|} \cdot \sum_{n}n|D_n|\cdot P_n.
\end{equation}

\subsection{Data Generation} \label{sec:DGP}
The simulation algorithm is designed to use 
a causal graph to determine the functional relation between a set of covariates
and a resulting simulated node (i.e. treatment assignment, outcome, and censoring decision).
Once this graph is defined, 
the data is used to calculate the values of the simulated nodes.

The causal graphs themselves are chosen randomly according to a set of parameters.  
Each such set of parameters defines a {\em data-generating process} (DGP),
and given a random seed, each DGP can provide a data instance pair of 
counter-factual results and factual results.

Parameters include the number of covariates controlling each simulated node 
and how many of them overlap (i.e. affect more than one node).
Specifically, a set of features is randomly chosen and connected to the outcome node;
another set of features is chosen and connected to the treatment assignment node;
and another set of features (possibly including the treatment assignment) is connected to the censoring node.
These nodes act in unison to determine the factual results that are reported.
Specifically, both counter-factual outcomes, censoring, and treatment assignment
are first calculated.
Then, if censoring is true then a value of {\tt NA} is reported as the observed outcome.
Otherwise, depending on the value of the treatment assignment,
the corresponding counter-factual outcome is reported.
To ensure that our simulation mimics a situation with no unmeasured confounders 
we do not use additional simulated data 
and share all the covariates the DGPs were based upon. 
Other parameters define treatment prevalence (i.e. percentage of treated); 
the amount of noise in the system (e.g. heterogeneity of treatment effect
\cite{velentgas2013developing});
and the degree of non-linearity between a simulated variable and it's causing covariates
(i.e. the degree of the polynomial, whether to use an exponential transform, etc.).

It should be noted that the code for simulating the data 
is currently not part of the benchmarking framework.
This is because the framework was chosen to be used in the causal inference challenge,
as part of the Atlantic Causal Inference Conference (ACIC2018).
Details on the challenge can be found at
\href{https://www.cmu.edu/acic2018/data-challenge}
{https://www.cmu.edu/acic2018/data-challenge}.

\section{Discussion and Conclusion}
We presented here a comprehensive benchamrking framework for the evaluation of methods 
for causal inference of treatment effects. 
The framework includes covariate data, simulated treatment assignment, 
simulated counter-factual outcomes, and code for evaluation of estimated effect. 
The framework is distributed freely using \emph{Apache 2.0} 
license on github.com.

We believe that, like in other fields of research, such benchmarking platforms
can allow a better comparison between methods, 
including the limitations and advantages of each method.
This will allow the community to jointly identify potential pitfalls and the areas most deserving effort.

We note also that there is no need to be confined to the data we provided, 
and individuals can still use the evaluation code with their own data.
Furthermore, since the data and code reside on github, 
important contributions can be suggested by the community. 
We encourage the community to work with us to contribute additional data-sets,
and welcome suggestions for more benchmarking metrics.
Such suggestions may be incorporated into the framework and be available to the community 
as a new version of the benchamrking framework.

\end{document}